\documentclass[aps,prd,showpacs,amssymb,amsmath,amsfonts,superscriptaddress,
twocolumn,nofootinbib,floatfix,preprintnumbers,altaffilletter]{revtex4}
\newif\ifpdf\ifx\pdfoutput\undefined\pdffalse\else\pdfoutput=1\pdftrue\fi
       \newcommand{\pdfgraphics}{\ifpdf\DeclareGraphicsExtensions{.pdf,.jpg}\else\fi}
\usepackage{graphicx}
\usepackage{hyperref}
\usepackage{amssymb}
\usepackage{amsmath}
\usepackage{longtable}
\usepackage{rotating}
\usepackage{color}
\usepackage{epsfig}
\usepackage{epsf}
\usepackage{fancyhdr}

\newcommand{\ee}[1]{\!\times\!10^{#1}}

\newcommand{\avec}{\mbox{\boldmath$a$}}
\begin{document}
\pdfgraphics \pagestyle{fancy}
\rhead[]{} \lhead[]{}
\title{
Limits on gravitational wave emission from selected pulsars using LIGO data\\
    }
%
% This list was generated on 10 Sept. 2004 A4 DCC LIGO - T040086-11-Z
\newcommand*{\AG}{Albert-Einstein-Institut, Max-Planck-Institut f\"ur Gravitationsphysik, D-14476 Golm, Germany}
\affiliation{\AG}
\newcommand*{\AH}{Albert-Einstein-Institut, Max-Planck-Institut f\"ur Gravitationsphysik, D-30167 Hannover, Germany}
\affiliation{\AH}
\newcommand*{\AN}{Australian National University, Canberra, 0200, Australia}
\affiliation{\AN}
\newcommand*{\CH}{California Institute of Technology, Pasadena, CA  91125, USA}
\affiliation{\CH}
\newcommand*{\DO}{California State University Dominguez Hills, Carson, CA  90747, USA}
\affiliation{\DO}
\newcommand*{\CA}{Caltech-CaRT, Pasadena, CA  91125, USA}
\affiliation{\CA}
\newcommand*{\CU}{Cardiff University, Cardiff, CF2 3YB, United Kingdom}
\affiliation{\CU}
\newcommand*{\CL}{Carleton College, Northfield, MN  55057, USA}
\affiliation{\CL}
\newcommand*{\FN}{Fermi National Accelerator Laboratory, Batavia, IL  60510, USA}
\affiliation{\FN}
\newcommand*{\HC}{Hobart and William Smith Colleges, Geneva, NY  14456, USA}
\affiliation{\HC}
\newcommand*{\IU}{Inter-University Centre for Astronomy  and Astrophysics, Pune - 411007, India}
\affiliation{\IU}
\newcommand*{\CT}{LIGO - California Institute of Technology, Pasadena, CA  91125, USA}
\affiliation{\CT}
\newcommand*{\LM}{LIGO - Massachusetts Institute of Technology, Cambridge, MA 02139, USA}
\affiliation{\LM}
\newcommand*{\LO}{LIGO Hanford Observatory, Richland, WA  99352, USA}
\affiliation{\LO}
\newcommand*{\LV}{LIGO Livingston Observatory, Livingston, LA  70754, USA}
\affiliation{\LV}
\newcommand*{\LU}{Louisiana State University, Baton Rouge, LA  70803, USA}
\affiliation{\LU}
\newcommand*{\LE}{Louisiana Tech University, Ruston, LA  71272, USA}
\affiliation{\LE}
\newcommand*{\LL}{Loyola University, New Orleans, LA 70118, USA}
\affiliation{\LL}
\newcommand*{\MP}{Max Planck Institut f\"ur Quantenoptik, D-85748, Garching, Germany}
\affiliation{\MP}
\newcommand*{\MS}{Moscow State University, Moscow, 119992, Russia}
\affiliation{\MS}
\newcommand*{\ND}{NASA/Goddard Space Flight Center, Greenbelt, MD  20771, USA}
\affiliation{\ND}
\newcommand*{\NA}{National Astronomical Observatory of Japan, Tokyo  181-8588, Japan}
\affiliation{\NA}
\newcommand*{\NO}{Northwestern University, Evanston, IL  60208, USA}
\affiliation{\NO}
\newcommand*{\SC}{Salish Kootenai College, Pablo, MT  59855, USA}
\affiliation{\SC}
\newcommand*{\SE}{Southeastern Louisiana University, Hammond, LA  70402, USA}
\affiliation{\SE}
\newcommand*{\SA}{Stanford University, Stanford, CA  94305, USA}
\affiliation{\SA}
\newcommand*{\SR}{Syracuse University, Syracuse, NY  13244, USA}
\affiliation{\SR}
\newcommand*{\PU}{The Pennsylvania State University, University Park, PA  16802, USA}
\affiliation{\PU}
\newcommand*{\TC}{The University of Texas at Brownsville and Texas Southmost College, Brownsville, TX  78520, USA}
\affiliation{\TC}
\newcommand*{\TR}{Trinity University, San Antonio, TX  78212, USA}
\affiliation{\TR}
\newcommand*{\HU}{Universit{\"a}t Hannover, D-30167 Hannover, Germany}
\affiliation{\HU}
\newcommand*{\BB}{Universitat de les Illes Balears, E-07122 Palma de Mallorca, Spain}
\affiliation{\BB}
\newcommand*{\BR}{University of Birmingham, Birmingham, B15 2TT, United Kingdom}
\affiliation{\BR}
\newcommand*{\FA}{University of Florida, Gainesville, FL  32611, USA}
\affiliation{\FA}
\newcommand*{\GU}{University of Glasgow, Glasgow, G12 8QQ, United Kingdom}
\affiliation{\GU}
\newcommand*{\MU}{University of Michigan, Ann Arbor, MI  48109, USA}
\affiliation{\MU}
\newcommand*{\OU}{University of Oregon, Eugene, OR  97403, USA}
\affiliation{\OU}
\newcommand*{\RO}{University of Rochester, Rochester, NY  14627, USA}
\affiliation{\RO}
\newcommand*{\UW}{University of Wisconsin-Milwaukee, Milwaukee, WI  53201, USA}
\affiliation{\UW}
\newcommand*{\WU}{Washington State University, Pullman, WA 99164, USA}
\affiliation{\WU}
\author{B.~Abbott}    \affiliation{\CT}
\author{R.~Abbott}    \affiliation{\LV}
\author{R.~Adhikari}    \affiliation{\LM}
\author{A.~Ageev}    \affiliation{\MS}  \affiliation{\SR}
\author{B.~Allen}    \affiliation{\UW}
\author{R.~Amin}    \affiliation{\FA}
\author{S.~B.~Anderson}    \affiliation{\CT}
\author{W.~G.~Anderson}    \affiliation{\TC}
\author{M.~Araya}    \affiliation{\CT}
\author{H.~Armandula}    \affiliation{\CT}
% ...
 \collaboration{The LIGO Scientific Collaboration, http://www.ligo.org}
 \noaffiliation
%
% This list was generated on 08 Sep 2004 - special CW author block
\newcommand*{\MA}{University of Manchester, Jodrell Bank Observatory, Macclesfield, Cheshire, SK11 9DL, United Kingdom}
%\affiliation{\MA}
%
\author{M.~Kramer}\affiliation{\MA}
\author{A.~G.~Lyne}\affiliation{\MA}
% \collaboration{The LIGO Scientific Collaboration, http://www.ligo.org}
% \noaffiliation
%
\date{\today}
\begin{abstract}
We place direct upper limits on the amplitude of gravitational
waves from 28 isolated radio pulsars by a coherent multi-detector
analysis of the data collected during the second science run of
the LIGO interferometric detectors. These are the first {\it
direct} upper limits for 26 of the 28 pulsars. We use coordinated
radio observations for the first time to build radio-guided phase
templates for the expected gravitational wave signals. The
unprecedented sensitivity of the detectors allow us to set strain
upper limits as low as a few times $10^{-24}$. These strain limits
translate into limits on the equatorial ellipticities of the
pulsars, which are smaller than $10^{-5}$ for the four closest
pulsars.
\end{abstract}
\pacs{04.80.Nn, 95.55.Ym, 97.60.Gb, 07.05.Kf}
%\preprint{LIGO-P040008-03-Z}
%
\maketitle
%
%%%%%%%%%%%%%%%%%%%%%%%%%%%%%%%%%%%%%%%%%%%%%%%%%%%%%%%%%%%%%%%%%%%
A worldwide effort is underway to detect gravitational waves (GWs)
and thus test a fundamental prediction of General Relativity. In
preparation for long-term operations, the LIGO and GEO experiments
conducted their first science run (S1) during 17 days in 2002. The
detectors and the analyses of the S1 data are described in Refs.
\cite{S1exp} and \cite{S1PulPaper}-\cite{S1StochPaper},
respectively. LIGO's second science run (S2) was carried out from
14 Feb to 14 April 2003, with dramatically improved sensitivity
compared to S1. During S2 the GEO detector was not operating.

A spinning neutron star is expected to emit GWs if it is not
perfectly symmetric about its rotation axis. The strain amplitude
$h_0$ of the emitted signal is proportional to the star's
deformation as measured by its ellipticity $\epsilon$ \cite{JKS}.
Using data from S2, this paper reports \emph{direct} observational
limits on the GW emission and corresponding ellipticities from the
28 most rapidly rotating isolated pulsars for which radio data is
complete enough to guide the phase of our filters with sufficient
precision. These are the first such limits for 26 of the pulsars.
We concentrate on isolated pulsars with known phase evolutions and
sky positions to ensure that our targeted search requires
relatively few unknown parameters.

\begin{figure}%[h]
  \begin{center}
    \includegraphics[width=\columnwidth,angle=0]{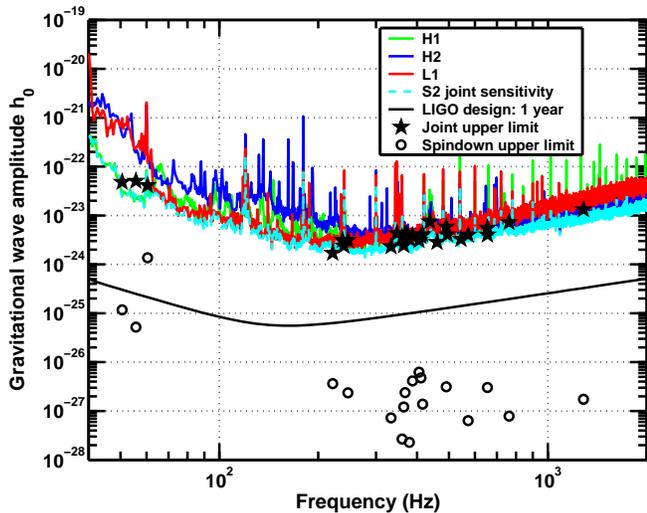}
  \end{center}
 \vspace*{-2.5em} \caption{Upper curves: $h_0$ amplitudes
detectable from a known generic source with a 1\% false alarm rate
and 10\% false dismissal rate, as given by Eq.~(2.2) in
\cite{S1PulPaper} for single detector analyses and for a joint
detector analysis. All the curves use typical S2 sensitivities and
observation times. H1 and H2 are the 4\,km-arm and the 2\,km-arm
detectors located in Hanford WA. L1 is the 4\,km-arm detector
situated in Livingston Parish LA. Lower curve: LIGO design
sensitivity for 1\,yr of data. Stars: upper limits found in this
paper for 28 known pulsars. Circles: spindown upper limits for the
pulsars with negative frequency derivative values if \emph{all}
the measured rotational energy loss were due to gravitational
waves and assuming a moment of inertia of $10^{45} \,{\rm g} \,
{\rm cm}^2$ .}\label{ifosens}
\end{figure}

The limits reported here are still well above the indirect limits
inferred from observed pulsar spindown, where available
(Fig.~\ref{ifosens}). However, fourteen of our pulsars are in
globular clusters, where local gravitational accelerations produce
Doppler effects that mask the intrinsic pulsar spindown, sometimes
even producing apparent spinup. For these pulsars our observations
therefore place the first limits that are inherently independent
of cluster dynamics, albeit at levels well above what one would
expect if all globular cluster pulsars are similar to field
pulsars.

Our most stringent ellipticity upper limit is $4.5 \times
10^{-6}$. While still above the maximum expected from conventional
models of nuclear matter, distortions of this size would be
permitted within at least one exotic theory of neutron star
structure \cite{bumps}.

\emph{Detectors.}---LIGO comprises three detectors. Each detector
is a power-recycled Michelson interferometer, with Fabry-Perot
cavities in the long arms. A passing GW produces a time-varying
differential strain in these arms, and the resulting differential
displacement of the cavity test mass mirrors is sensed
interferometrically. Two detectors, the 4\,km-arm H1 and the
2\,km-arm H2 detectors, are collocated in Hanford WA.  The
4\,km-arm L1 detector is situated in Livingston Parish LA.
Improvements in noise performance between S1 and S2 were
approximately an order of magnitude over a broad frequency range.
Modifications that were made between S1 and S2 to aid in noise
reduction and improve stability include: i) increased laser power
to reduce high-frequency noise, ii) better angular control of the
mirrors of the interferometer and iii) the use of lower noise
digital test mass suspension controllers in all detectors.

During S2, the LIGO detectors' noise performance in the band
40-2000\,Hz was better than any previous interferometer. The best
strain sensitivity, achieved by L1, was $\sim 3 \times
10^{-22}{\rm\,Hz}^{-1/2}$ near 200\,Hz (where it translates
through Eq.~(2.2) of \cite{S1PulPaper} into a detectable amplitude
for a continuous signal of about $3\times 10^{-24}$, as shown in
Fig.~\ref{ifosens}). The relative timing stability between the
interferometers was also significantly improved. Monitored with
GPS-synchronized clocks to be better than $10\,\mu {\rm{s}}$ over
S2, it allowed the coherent combination of the strain data of all
three detectors to form joint upper limits.

%\section{ANALYSIS METHOD}

\emph{Analysis method.}---In \cite{S1PulPaper} a search for
gravitational waves from the millisecond pulsar J1939$+$2134 using
S1 data was presented. In that work, two different data analysis
methods were used, one in the time domain and the other in the
frequency domain. Here we extend the former method
\cite{S1PulPaper, TDmethods} and apply it to 28 isolated pulsars.

Following \cite{S1PulPaper} we model the sources as non-precessing
triaxial neutron stars showing the same rotational phase evolution
as is present in the radio signal and perform a complex heterodyne
of the strain data from each detector at the instantaneous
frequency of the expected gravitational wave signal, which is
twice the observed radio rotation frequency. These data are then
down-sampled to $1/60$\,Hz and will be referred to as $B_k$. Any
gravitational signal in the data would show a residual time
evolution reflecting the antenna pattern of the detector, varying
over the day as the source moved through the pattern, but with a
functional form that depended on several other source-observer
parameters: the antenna responses to plus and cross polarisations,
the amplitude of the gravitational wave $h_0$, the angle between
the line-of-sight to the pulsar and its spin axis $\iota$, the
polarisation angle of the gravitational radiation $\psi$ (all
defined in \cite{JKS}) and the phase $\phi_0$ of the gravitational
wave signal at some fiducial time $t_0$. Let $\avec$ be a vector
in parameter space with components $(h_0, \iota, \psi, \phi_0)$.

The analysis proceeds by determining the posterior probability
distribution function (pdf) of $\avec$ given the data $B_k$ and
the signal model:
 \begin{equation}
p(\avec|\{B_k\})\propto p(\avec)p(\{B_k\}|\avec),
\label{bayes}
 \end{equation}
where $p(\{B_k\}|\avec)$ is the likelihood and $p(\avec)$ the
prior pdf we assign to the model parameters. We have used a
uniform prior for $\cos\iota$, $\phi_0$, $\psi$ and $h_0$
($h_0>0$), in common with \cite{S1PulPaper}. A uniform prior for
$h_0$ has been chosen for its simplicity and so that our results
can readily be compared with other observations. This prior favors
high values of $h_0$ (which comprise the majority of the parameter
space) and therefore generates a somewhat conservative upper limit
for its value. Indeed the reader might prefer to regard our
resulting posterior pdfs for $h_0$ as marginalised likelihoods
rather than probabilities for $h_0$  --- these are functionally
identical using our priors.

As in \cite{S1PulPaper} we use a Gaussian joint likelihood for
$p(\{B_k\}|\avec)$. In \cite{S1PulPaper} the S1 noise floor was
estimated over a 60\,s period from a 4\,Hz band about the expected
signal frequency. This gave a reliable point estimate for the
noise level but was sensitive to spectral contamination within the
band, as demonstrated in the analysis of the GEO S1 data. In this
paper we exploit the improved stationarity of the instruments that
make it reasonable to assume the noise floor is constant over
periods of 30\,min \cite{TDmethods}. In addition we restrict the
bandwidth to $1/60$\,Hz, which makes it possible to search for
signals from pulsars at frequencies close to strong spectral
disturbances. However, the noise level now determined is less
certain as the estimate relies on fewer data. We take account of
this increased uncertainty by explicitly marginalising with a
Jeffreys prior over the constant but unknown noise level for each
30\,min period of data \cite{Bretthorst}. The likelihood for this
analysis is then the combined likelihood for all the 30\,min
stretches of data, labeled by $j$ in Eq.~(\ref{eq:likelihood}),
taken as independent:
\begin{eqnarray}
 p(\{B_k\}|\avec) &\propto&  \prod_j p(\{B_k\}_j|\avec),
\label{eq:likelihood}
\\
 p(\{B_k\}_j|\avec) &\propto&
 \left(\sum_{k=k_{1(j)}}^{k_{2(j)}}|B_k-y_k|^2\right)^{-m},
\label{student}
\end{eqnarray}
where $y_k$ is the signal model given by Eq.~(4.10) in
\cite{S1PulPaper} and $m = k_{2(j)}-k_{1(j)}+1 = 30$ is the number
of $B_k$ data points in a 30\,min segment.

In principle the period over which the data are assumed stationary
need not be fixed, and can be adjusted dynamically to reflect
instrumental performance over the run. We have limited our
analysis to continuous 30\,min stretches of data, which included
more than 88\% of the S2 science data set. Inclusion of shorter
data sections would at best have resulted in a $\sim 6\%$
improvement on the strain upper limits reported here.

\begin{figure}%[h]
  \begin{center}
    \includegraphics[width=\columnwidth,angle=0]{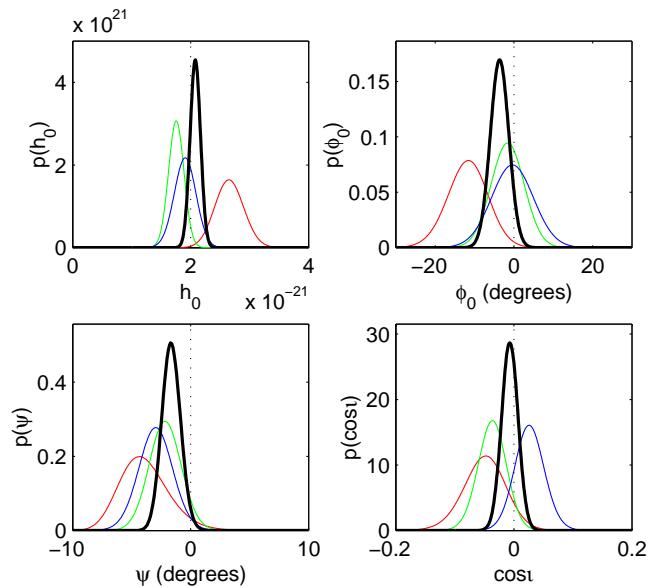}
  \end{center}
 \vspace*{-2em}
\caption{Parameters of the artificial pulsar P1, recovered from
12\,h of strain data from the Hanford and Livingston
interferometers.  The results are displayed as marginal pdfs for
each of the four signal parameters. The vertical dotted lines show
the values used to generate the signal, the colored lines show the
results from the individual detectors (H1 green, H2 blue, L1 red),
and the black lines show the joint result from combining
coherently data from all three.} \label{injections}
\end{figure}

\emph{Validation by hardware injections.}---All the software used
in this analysis is available in the LAL and LALapps CVS
repositories (www.lsc-group.phys.uwm.edu/daswg/projects/lal.html)
with the tag ``pulgroup\_paper\_s2tds''. It was validated by
checking its performance on fake pulsar signals injected in
artificial and real detector noise, both in software
(\cite{S1PulPaper}) and in hardware. In particular, two artificial
signals (P1, P2) were injected into all three detectors by
modulating the mirror positions via the actuation control signals
with the strain signal we should expect from a hypothetical
pulsar. These injections were designed to give an end-to-end
validation of the search pipeline starting from as far up the
observing chain as possible.

The pulsar signals were injected for 12\,h at frequencies of
1\,279.123\,Hz (P1) and 1\,288.901\,Hz (P2) with frequency
derivatives of zero and $-10^{-8}$\,Hz\,s$^{-1}$ respectively, and
strain amplitudes of $2\times10^{-21}$.  This gives
signal-to-noise ratios (as defined by Eq.~(79) of \cite{JKS}) of
$26$ and $40$ for P1 in H1 and L1 respectively and of $38$ and
$34$ for P2. The signals were modulated and Doppler shifted to
simulate sources at fixed positions on the sky with $\psi=0$,
$\cos\iota=0$ and $\phi_0=0$. To illustrate, posterior pdfs for
the recovered P1 signal are shown in Fig.~\ref{injections}. The
results derived from the different detectors are in broad
statistical agreement,  confirming that the relative calibrations
are consistent and that the assessments of uncertainty (expressed
in the posterior widths) are reasonable. Results for P2 were very
similar to these.

The phase stability of the detectors in S2 allowed us to implement
a \emph{joint} coherent analysis based on data from all three
participating instruments. This technique was noted in
\cite{S1PulPaper}, but could not be performed on the S1 data
because of timing uncertainties that existed when those
observations were performed. The black lines in
Fig.~\ref{injections} show marginalisations of the joint posterior
from H1, H2 and L1, i.e.,
\begin{equation}
 p(\avec|\text{H1,\,H2,\,L1})\propto p(\avec)  \,
 p(\text{H1}|\avec)\, p(\text{H2}|\avec) \,
 p(\text{L1}|\avec).
\end{equation}
In an ideal case of three detectors of similar sensitivities and
operational times these coherent results would be approximately
$\sqrt{3}$ times tighter than the individual results. The
posteriors for $\phi_0$ clearly highlight the relative coherence
between the instruments and verify that similar joint methods can
be used to set upper limits on our target pulsars.

\emph{Results.}---From the ATNF pulsar catalogue ({\tt
www.atnf.csiro.au/research/pulsar/psrcat/}) we selected 28
isolated pulsars with rotational frequencies greater than 20\,Hz
and for which sufficiently good timing data were available
(Table~I). For 18 of these, we obtained updated timing solutions
from regular timing observations made at the Jodrell Bank
Observatory using the Lovell and the Parkes telescopes, adjusted
for a reference epoch centred on the epoch of the S2 run (starred
pulsars in Table~I). Details of the techniques that were used to
do this can be found in \cite{Timings}. We also checked that none
of these pulsars exhibited a glitch during this period.

The list includes globular cluster pulsars (including isolated
pulsars in 47~Tuc and NGC6752), the S1 target millisecond pulsar
(J1939+2134) and the Crab pulsar (B0531+21). Although Table~I only
shows approximate pulsar frequencies and frequency derivatives,
further phase corrections were made for pulsars with measured
second derivatives of frequency. Timing solutions for the Crab
were taken from the Jodrell Bank online ephemeris
\cite{CrabEphem}, and adjustments were made to its phase over the
period of S2 using the method of \cite{PW}.
\begin{table}
%\begin{center}
\tabcolsep 3pt
\begin{tabular}{lrrcc}
                 & spin~~   &       spindown    &$h_0^{95\%}$& $\epsilon$ \\
~~~~pulsar       & $f$ (Hz) &$\dot{f}$ (Hz\,s$^{-1}$)&$/10^{-24}$& $/10^{-5}$ \\ \hline
 B0021$-$72C$^*$ &   173.71 &   $+1.50\ee{-15}$ &   $4.3$ & $16 $   \\
 B0021$-$72D$^*$ &   186.65 &   $+1.19\ee{-16}$ &   $4.1$ & $14  $  \\
 B0021$-$72F$^*$ &   381.16 &   $-9.37\ee{-15}$ &   $7.2$ & $5.7 $  \\
 B0021$-$72G$^*$ &   247.50 &   $+2.58\ee{-15}$ &   $4.1$ & $7.5 $  \\
 B0021$-$72L$^*$ &   230.09 &   $+6.46\ee{-15}$ &   $2.9$ & $6.1 $  \\
 B0021$-$72M$^*$ &   271.99 &   $+2.84\ee{-15}$ &   $3.3$ & $5.0 $  \\
 B0021$-$72N$^*$ &   327.44 &   $+2.34\ee{-15}$ &   $4.0$ & $4.3 $  \\
 J0030+0451      &   205.53 &   $-4.20\ee{-16}$ &   $3.8$ & $0.48$  \\
 B0531+21$^*$    &    29.81 &   $-3.74\ee{-10}$ &   $41 $ & $2\,100$\\
 J0711$-$6830    &   182.12 &   $-4.94\ee{-16}$ &   $2.4$ & $1.8 $  \\
 J1024$-$0719$^*$&   193.72 &   $-6.95\ee{-16}$ &   $3.9$ & $0.86$  \\
 B1516+02A       &   180.06 &   $-1.34\ee{-15}$ &   $3.6$ & $21 $   \\
 J1629$-$6902    &   166.65 &   $-2.78\ee{-16}$ &   $2.3$ & $2.7 $  \\
 J1721$-$2457    &   285.99 &   $-4.80\ee{-16}$ &   $4.0$ & $1.8 $  \\
 J1730$-$2304$^*$&   123.11 &   $-3.06\ee{-16}$ &   $3.1$ & $2.5 $  \\
 J1744$-$1134$^*$&   245.43 &   $-5.40\ee{-16}$ &   $5.9$ & $0.83$  \\
 J1748$-$2446C   &   118.54 &   $+8.52\ee{-15}$ &   $3.1$ & $24  $  \\
 B1820$-$30A$^*$ &   183.82 &   $-1.14\ee{-13}$ &   $4.2$ & $24 $   \\
 B1821$-$24$^*$  &   327.41 &   $-1.74\ee{-13}$ &   $5.6$ & $7.1$   \\
 J1910$-$5959B   &   119.65 &   $+1.14\ee{-14}$ &   $2.4$ & $8.5 $  \\
 J1910$-$5959C   &   189.49 &   $-7.90\ee{-17}$ &   $3.3$ & $4.7 $  \\
 J1910$-$5959D   &   110.68 &   $-1.18\ee{-14}$ &   $1.7$ & $7.2 $  \\
 J1910$-$5959E   &   218.73 &   $+2.09\ee{-14}$ &   $7.5$ & $7.9 $  \\
 J1913+1011$^*$  &   27.85  &   $-2.61\ee{-12}$ &   $51 $ & $6\,900$\\
 J1939+2134$^*$  &   641.93 &   $-4.33\ee{-14}$ &   $13 $ & $2.7 $  \\
 B1951+32$^*$    &    25.30 &   $-3.74\ee{-12}$ &   $48 $ & $4\,400$\\
 J2124$-$3358$^*$&   202.79 &   $-8.45\ee{-16}$ &   $3.1$ & $0.45$  \\
 J2322+2057$^*$  &   207.97 &   $-4.20\ee{-16}$ &   $4.1$ & $1.8 $  \\
  \hline
\end{tabular}
\caption{The 28 pulsars targeted in the S2 run, with approximate
spin parameters. Pulsars for which radio timing data were taken
over the S2 period are starred (*). The right-hand two columns
show the 95\% upper limit on $h_0$, based on a coherent analysis
using all the S2 data, and corresponding ellipticity values
($\epsilon$, see text). These upper limit values do {\it{not}}
include the uncertainties due to calibration and to pulsar timing
accuracy,  which are discussed in the text, nor uncertainties in
the pulsar's distance, $r$.}
%\end{center}
\end{table}

The analysis used 910 hours of data from H1, 691 hours from H2,
and 342 hours from L1. There was no evidence of strong spectral
contamination in any of the bands investigated, such as might be
caused by an instrumental feature or a potentially detectable
pulsar signal.  A strong gravitational signal would generate a
parameter pdf prominently peaked off zero with respect to its
width, as for the  hardware injections. Such a pdf would trigger a
more detailed investigation of the pulsar in question. No such
triggers occurred in the analysis of these data, and we therefore
present upper limits.

The upper limits are presented as the value of $h_0$ bounding 95\%
of the cumulative probability of the marginalised strain pdf from
$h_0=0$.  The joint upper limit $h_0^{95\%}$ therefore satisfies
 \begin{equation}
 0.95 = \int_{h_0=0}^{h_0^{95\%}}{\rm d}h_0 \iiint
 p(\avec|\text{H1,\,H2,\,L1}){\rm d}\iota \,{\rm d}\psi\,{\rm d}\phi_0,
 \end{equation}
consistent with \cite{S1PulPaper}.  The uncertainty in the noise
floor estimate is already included, as outlined above.

The remaining uncertainties in the upper limit values of Table~I
stem from the calibration of the instrument and from the accuracy
of the pulsar timing models.  For L1 and H2, the amplitude
calibration uncertainties are conservatively estimated to be 10\%
and 8\%, respectively.  For H1, the maximum calibration
uncertainty is 18\%, with typical values at the 6\% level.  Phase
calibration uncertainties are negligible in comparison: less than
$10^\circ$ in all detectors. Biases due to pulsar timing errors
are estimated to be 3\% or less for J0030+0451, and 1\% or less
for the remaining pulsars (see \cite{S1PulPaper} for a discussion
of the effect of these uncertainties).

\emph{Discussion.}---The improved sensitivity of the LIGO
interferometers is clear from the strain upper limit for
PSR~J1939+2134, which is more than a factor of ten lower than was
achieved with the S1 data \cite{S1PulPaper}. In this analysis the
lowest limit is achieved for PSR~J1910$-$5959D at the level of
$1.7\times10^{-24}$, largely reflecting the lower noise floor
around 200\,Hz.

Table~I also gives approximate limits to the ellipticities \cite{JKS} of
these pulsars from the simple quadrupole model
\begin{equation}
\epsilon  \simeq 0.237 \frac{h_0}{10^{-24}} \frac{r}{1\,{\rm kpc}}
\frac{1\,{\rm Hz}^2}{f^2} \frac{10^{45}{\rm \,g\,cm^2}}{I_{zz}}
\label{eq:epsilon}
\end{equation}
where $r$ is the pulsar's distance, which we take as the
dispersion measure distance using the model of Taylor and Cordes
\cite{TC93}, and $I_{zz}$ its principal moment of inertia about
the rotation axis, which we take as $10^{45}$\,g\,cm$^2$.

As expected, none of these upper limits improves on those inferred
from simple arguments based on the gravitational luminosities
achievable from the observed loss of pulsar rotational kinetic
energy.  However, as discussed in the introduction, for pulsars in
globular clusters such arguments are complicated by cluster
dynamics, which the direct limits presented here avoid.

The result for the Crab pulsar (B0531+21) is within a factor of
about 30 of the spindown limit and over an order of magnitude
better than the previous direct upper limit of \cite{CrabUL}. The
equatorial ellipticities of the four closest pulsars (J0030+0451,
J2124+3358, J1024$-$0719, and J1744$-$1134) are constrained to be
less than $10^{-5}$.

Once the detectors operate at design sensitivity for a year, the
observational upper limits will improve by more than an order of
magnitude. The present analysis will also be extended to include
pulsars in binary systems, significantly increasing the population
of objects under inspection.

\emph{Acknowledgments.}---The authors gratefully acknowledge the
support of the United States National Science Foundation for the
construction and operation of the LIGO Laboratory and the Particle
Physics and Astronomy Research Council of the United Kingdom, the
Max-Planck-Society and the State of Niedersachsen/Germany for
support of the construction and operation of the GEO600 detector.
The authors also gratefully acknowledge the support of the
research by these agencies and by the Australian Research Council,
the Natural Sciences and Engineering Research Council of Canada,
the Council of Scientific and Industrial Research of India, the
Department of Science and Technology of India, the Spanish
Ministerio de Ciencia y Tecnologia, the John Simon Guggenheim
Foundation, the Leverhulme Trust, the David and Lucile Packard
Foundation, the Research Corporation, and the Alfred P.~Sloan
Foundation. This document has been assigned LIGO Laboratory
document number LIGO-P040008-B-Z.
%%%%%%%%%%%%%%%%%%%%%%%%%%%%%%%%%%%%%%%%%%%%%%%%%%%%%%%%%%%%%%%%%%%%%%
% references

%
\end{document}